\documentclass[]{spie}  

\usepackage{amsmath,amsfonts,amssymb}
\usepackage{graphicx}

\title{The Ingot WFS ON an ELT-like telescope: the project and simulations}
\author[a,b]{Portaluri E.} 
\author[c,b]{Di Filippo S.} 
\author[c,b]{Viotto V. }
\author[d,b]{Ragazzoni R.} 
\author[c,b]{Arcidiacono C.} 
\author[c,b]{Greggio D.}
\author[c,b]{Radhakrishnan Santhakumari  K. K.}
\author[c,b]{Bergomi M.} 
\author[c,b]{Marafatto L. }
\author[c,b]{Dima M.,}
\author[c,b]{Farinato J.} 
\author[c,b]{Magrin D.} 
\author[a,b]{Di Rico G.} 
\author[e,b]{Centrone M.}
\author[f,b]{Bonaccini D.}

\affil[a]{INAF - Osservatorio Astronomico d'Abruzzo, Via Mentore Maggini, I-64100 Teramo, Italy}
\affil[b]{ADONI, Laboratorio Nazionale di Ottica Adattiva, Italy}
\affil[c]{INAF - Osservatorio Astronomico di Padova, Vicolo dell’Osservatorio 5, I-35122 Padova, Italy }
\affil[d]{Dipartimento di Fisica e Astronomia, Università degli Studi di Padova, Vicolo dell’Osservatorio 3, I-35122 Padova, Italy}
\affil[e]{INAF - Osservatorio Astronomico Roma, Via Frascati 33, I-00078, Monte Porzio Catone (RM)}
\affil[f]{European Southern Observatory, Karl-Schwarzschildstr. 2, 85748 Garching, Germany}

\authorinfo{Further author information: (Send correspondence to EP.)\\EP: E-mail: elisa.portaluri$@$inaf.it}

\begin{document}
\maketitle

\begin{abstract}
  The Ingot WFS represents an innovative and indispensable class of sensors conceived to overcome some limitations due to the LGSs geometry, which is significantly different from the point-object originated by a NGS. Here we overview the project, aiming at investigating the performance of an ELT-like telescope equipped with the Ingot WFS, facing different aspects of the program: the needs for numerical simulations and laboratory experiments, the prototype and, finally the future plan for the verification on sky. 
\end{abstract}


\section{INTRODUCTION}
\label{sec:intro}  
The next generation of extremely large telescopes (ELTs, i.e. the European Extremely Large Telescope \cite{Gilmozzi2007}, ELT, and the Giant Magellan Telescope \cite{Johns2008}) MCAO modules are going to be fed by artificial sources, fired by launchers located at the side of the telescope that excites the Sodium layer. Being generated at around 90 km from the ground, e.g. at a finite distance, and on a stratum that has a variable thickness, these sources are not point-like objects, but something similar to "cigars". 
The Ingot wavefront sensor (I-WFS \cite{Ragazzoni2017,Ragazzoni2018, Ragazzoni2019}), shaped like an prism, is designed to cope with this typical elongation and thus overcoming such a limitation in order to improve the Adaptive Optics (AO) correction\cite{Fried1995,Pfrommer2009}. 

The project here described is an $R\&D$ study aimed at the investigation of the feasibility of this innovative and ground-breaking wavefront sensor, reminding that its advancement can be considered as a risk mitigator for the AO systems of the future ELTs, whose baselines foresee only the use of “conventional” sensors with such elongated sources.
It is developed by a passionate group of young researchers ($\sim$15), belonging to INAF institutes of Padova, Teramo and Rome and accounts for numerical simulations, lab experiments, the prototype and, finally, the verification on sky at the CANARY facility of the William Herschel Telescope (WHT) in La Palma.

Here, we present the current status of the project, reporting the operations we are setting out to explore how the sensor can be implemented on large telescopes to fully exploit their extraordinary sensitivity and resolution capabilities.
In particular, we focus on the description of the overall architecture and results of ongoing activities: the simulations we set out to investigate the performance of the I-WFS on an ELT-like telescope (Section~\ref{sec:sim}) and the laboratory tests we made to study the response of the sensor (Section~\ref{sec:lab}). Then we discuss the plan for the future, accounting for the test of the sensor on the Laser Laboratory at the Astronomical Observatory of Rome and, finally, the verification on sky (Section~\ref{sec:sky}) at the ESO Wendelstein Laser Guide Star Unit (WLGSU) at WHT. 

\section{Numerical Simulations}
\label{sec:sim}
We developed a simulation tool to evaluate the impact of the ingot WFS and test its performance. 

The peculiarity of the project lies precisely in the different approach we use to face an old problem: for the first time we have to deal with E2E simulations that would consider a 3D space, as the Fourier (2D) method, commonly used, would not enhance the advantages of the sensor, designed to couple the geometric characteristics of the LGS~\cite{Viotto2018,Viotto2019}. 

\begin{figure} [hb]
   \begin{center}
   \begin{tabular}{c} 
   \includegraphics[height=9cm]{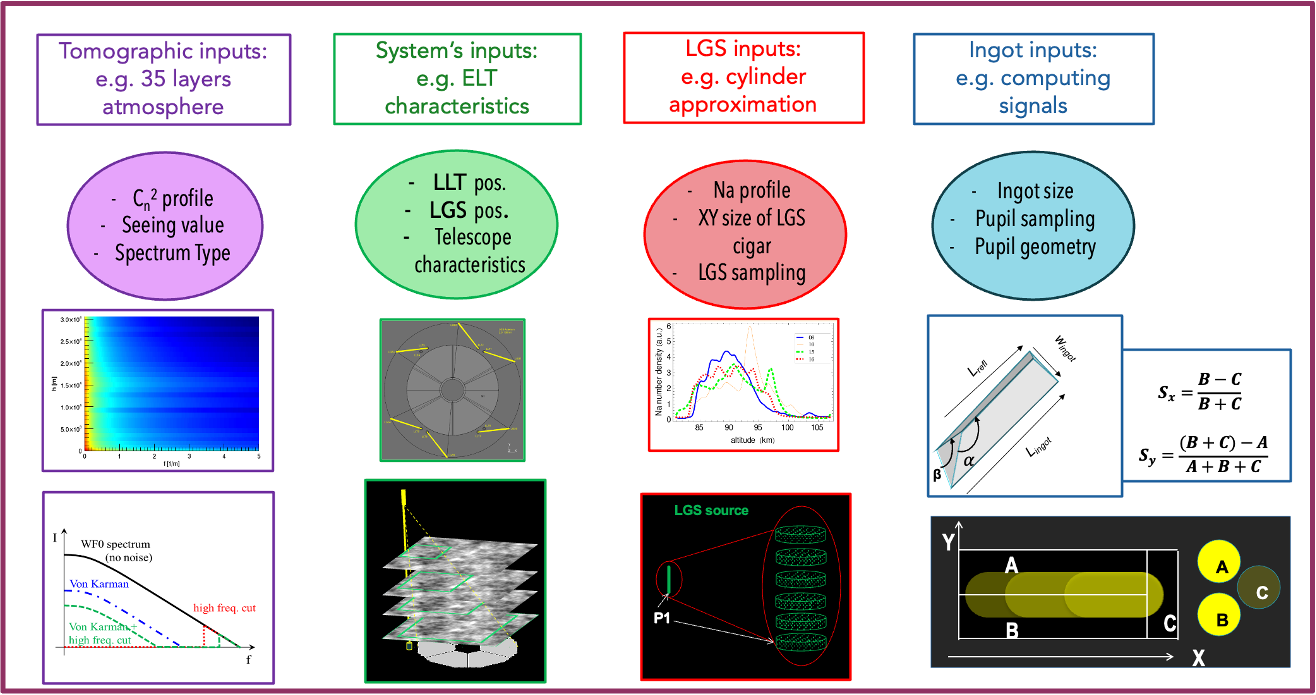}
   \end{tabular}
   \end{center}
   \caption[example] 
   { \label{fig:simulator} List of the 4 main categories that the software accounts for: tomography, system, LGS and ingot geometry. For each of them some examples of inputs are listed and shown with a figure or plot.}
   \end{figure}

The tool accepts a list of input parameters that describe the configuration whose final performance the user wants to estimate. 
These parameters include atmospheric assumptions, telescope and system characteristics, and purely ingot aspects, as shown in Figure~\ref{fig:simulator} and are fully described in previous papers~\cite{Portaluri2019, 2020SPIEElisa}. Therefore, assuming a given geometry of the system and an appropriate LGS asterism, accounting for the cone effect and for an input atmosphere, the simulator will retrieve the incoming wavefront. It is then translated into the focal plane, through the ingot, to the pupil plane, where the 3 pupils are re-imaged and then used to compute signals~\cite{2020SPIEElisa}, calculated with different options.
   
The software works in a closed-loop fashion, aiming at the reconstruction of the incoming turbulent wavefront using a modal approach. In particular, we use a Zernike modal basis to compute the interaction matrix and the relative reconstructor, which is calculated and then used to retrieve the output wavefront from signals.
The outcome is the achievable SR, which can be compared with the one expected for the next generation of instruments in the same conditions and used to quantify the gain reachable in many cases, also considering the science applications.

At the moment, we are working in a "comfortable zone", using a frozen atmosphere and first 100 Zernike modes, performing first loops in low-resolution mode. However, the first results are encouraging: we were able to properly close the loop and obtain very good performance, reconstructing the incoming wavefront with an error less than 2\%\cite{2020SPIEElisa}. 

A great effort and a lot of time time have been dedicated to overcome computational and memory issues in order to work (in the future) in the high-resolution regime (as an example, the high pupil sampling is a bottle neck) and, currently, we are ready to improve the description of the performance improving the input parameters as resolution, sampling, number of modes and dynamical disturbance and, at the end, to compare the results with those obtained using a SH-WFS.

Then, the plan is to explore the range of parameters to find which one mostly influences the response of the sensor, then to reproduce:
\begin{itemize}
   \item the laboratory configuration in order to compare the results (see Section~\ref{sec:lab})
    \item the ELT configuration as much as possible in order to compare the prediction of the AO correction, with the outcome of the I-WFS.
\end{itemize}

As the simulations are proceeding in parallel with first experiments on the laboratory~\cite{2020SPIEKalyan, DiFilippo2022} (Section~\ref{sec:lab}) and also with tests in the Fourier environment~\cite{Arcidiacono2020}, at the end we will have a complete view of the response of the sensor. Their comparison will be necessary in order to properly evaluate the positive impact of such new sensor on an AO system, going towards the possibility to test it on sky (Section~\ref{sec:sky}).

\section{Laboratory Experiments}
\label{sec:lab}
Laboratory experiments are required in order to have a hint of the effectiveness of the sensor, and they have to be close as much as possible to reality to properly couple the geometric characteristics of the LGS to the sensor itself: this translates on the fact that for the first time we had topay close attention to the 3rd dimension. 

In short, we dedicated 2 setups to investigate different features of the I-WFS, performing:
\begin{itemize}
    \item  a test in a quasi-real Adaptive Optics Closed Loop scenario at the LOOPS bench at LAM~\cite{Arcidiacono2020}
    \item  a test of the alignment of the system and measurements of low-order static aberrations in an Open Loop scenario~\cite{DiFilippo2019, 2020SPIEKalyan, DiFilippo2022}. 
\end{itemize}

\begin{figure} [h]
   \begin{center}
   \begin{tabular}{c} 
   \includegraphics[height=7cm]{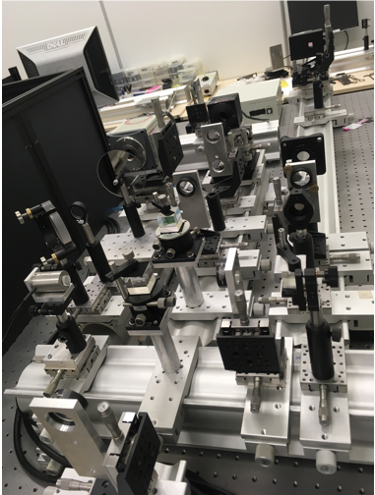}
   \end{tabular}
   \end{center}
   \caption[example] 
   { \label{fig:LAM} Ingot setup at LAM. For more details see Arcidiacono et al. 2020.}
   \end{figure}

In the first case, the I-WFS was approximated with a set of 2D phase masks and tested using different formulas for calculation of signals and configuration of the LGS elongation (Figure~\ref{fig:LAM}). The results were compared with the ones obtained with a 2D phase mask pyramid-like sensor and show the advantage of the I-WFS in the most elongated configuration~\cite{Arcidiacono2020}.
   
On the opposite side, following the second approach, we built an optical setup mimicking the ELT characteristics at the INAF-Observatory of Padova to test the I-WFS and its alignment procedure using an OLED screen to reproduce the LGS source, which is elongated along the optical axis, coupled with a real ingot prism.

Several simulations and tests were required to align the ingot prism to the LGS source, considering its 6 degrees of freedom. The procedure is now definitive, its working and efficacy is described in a previous paper~\cite{2020SPIEKalyan} and now it is fully automatized~\cite{DiFilippo2022} .

With this configuration we performed:
\begin{itemize}
    \item Movements simulation analysis
 \item Sensitivity analysis
 \item Definition and automatization of the alignment procedures
\end{itemize}

\begin{figure} [h]
   \begin{center}
   \begin{tabular}{c} 
   \includegraphics[height=3.4cm]{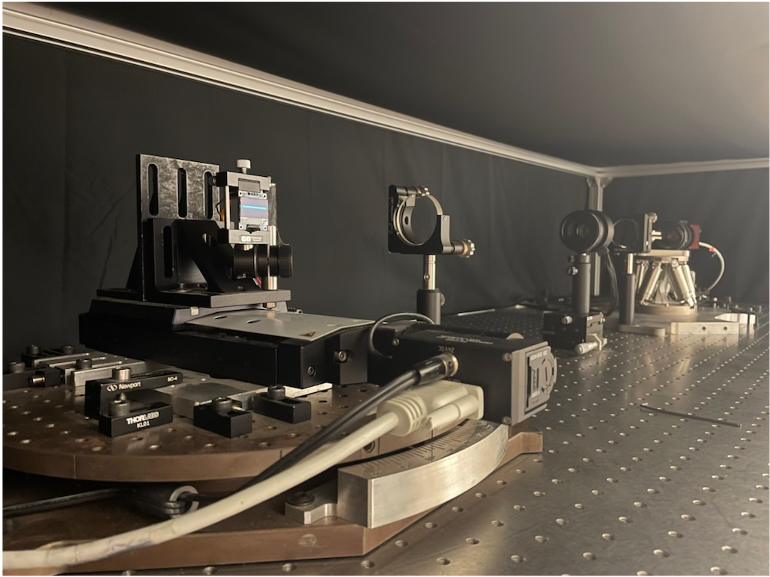}
   \includegraphics[height=3.4cm]{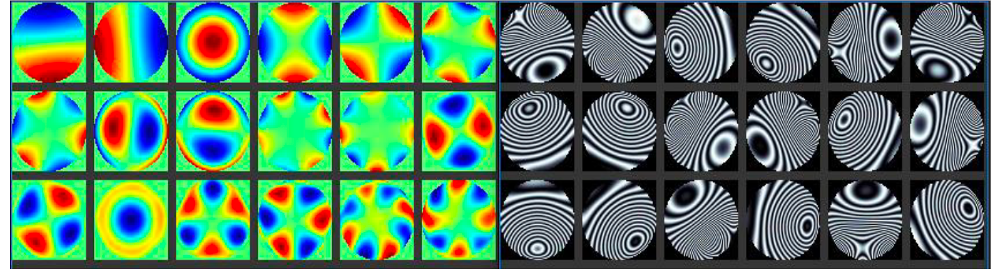}
   \end{tabular}
   \end{center}
   \caption[example] 
   { \label{fig:DL} Left-hand panel: Ingot setup at INAF-Padova, including the DL. central Right-hand panel: aberration modes produced by the 18 actuators of the DL and corresponding shapes of the actuators. }
   \end{figure}

Currently we positioned a deformable lens (DL, aperture= 25mm, n. actuators=18) in the pupil plane in order to apply low order aberrations to the wavefront (Figure~\ref{fig:DL}).
In this way, we can apply a single value or combination of modes and actuators movements using our Python/Matlab procedure, which controls the system, to create known aberrations\cite{DiFilippo2022}.

However, the laboratory setup is evolving in time to accommodate new components: for example in the future we plan to use in the future a deformable mirror for the close-loop operations and eventually a "classical" WF sensor (a pyramid and/or a Shack-Hartmann) to compare the performances. 
As a further step, we plan to mimic as much as possible the real scenario by the introduction of the source profile, which takes into account the temporal and spatial variability of the Sodium layer.
And finally, our goal is to reach an advanced phase (in the setup configuration and in the learned procedure) to make the laboratory results comparable the I-WFS performance obtained with the end-to-end simulations\cite{2020SPIEElisa}.
   
\section{Future Prototype and Verification}
\label{sec:sky}

As already highlighted, the I-WFS can be considered in the strategy of risk mitigation for the next generation of ELTs equipped with LGS launchers (located at the side of the telescopes.)
However, in order to demonstrate its feasibility and gain, we have to go in the direction of prototyping.

Once available, first, we will test the prototype and verify the performance on a stand-alone opto-mechanic system in the Laser Laboratory at INAF-OAR, provided by a LGS simulator and possibly a turbulence generator.
\begin{figure} [ht]
   \begin{center}
   \begin{tabular}{c} 
   \includegraphics[height=4.5cm]{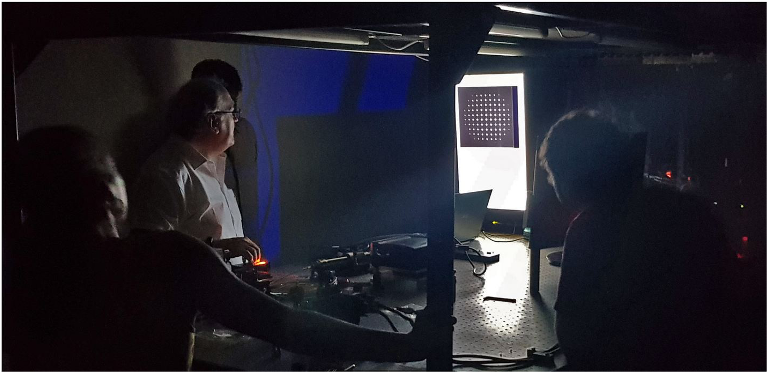}
   \includegraphics[height=4.5cm]{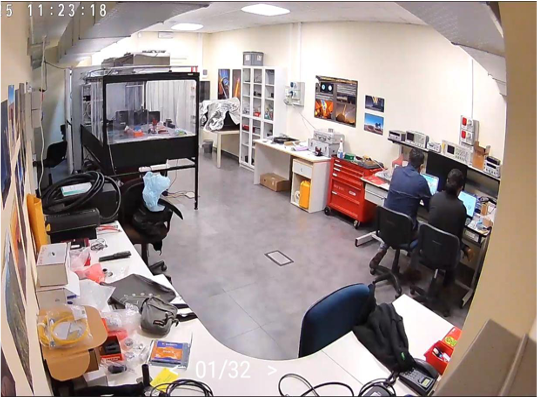}
   \end{tabular}
   \end{center}
   \caption[example] 
   { \label{fig:Rome} Laser Laboratory at INAF-Rome. }
   \end{figure} 
Then we plan to study the real working instrument and participate to the next year call at ESO Wendelstein Laser Guide Star Unit (WLGSU) at WHT, adapting interfaces and organizing the on-sky verification.
The WLGSU is located 40 m from the WHT,  and is used together with CANARY adaptive optics system on the WHT: a Sodium laser beam is launched from the WLGSU  as part of laser guide star field tests for validating ELT laser guide star baseline performance. 

\begin{figure} [ht]
   \begin{center}
   \begin{tabular}{c} 
   \includegraphics[height=6cm]{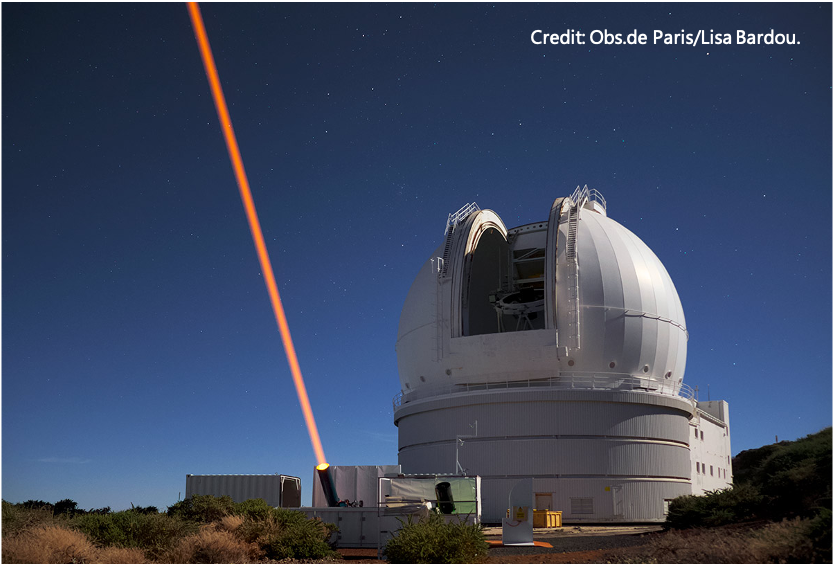}
   \end{tabular}
   \end{center}
   \caption[example] 
   { \label{fig:Rome} WHT with the WLGSU facility. Credit: Obs.de Paris$/$Lisa Bardou. }
   \end{figure} 

\section{Status of the project and conclusions}

The Ingot $R\&D$ project has an ambitious goal: to develop a new type of sensor that allows to optimize the performance of the AO systems of the new generation telescopes, overcoming some well-known limits, related to the use of LGSs. 

To achieve this goal, several phases of experimentation and testing are required (in laboratory and through simulations), all well defined, of considerable technological and theoretical effort. 
In fact, we remark here that the peculiarity of this study lies precisely in the different approach to manage a well-known problem:  for the first time we have to carry out 3D tests in the laboratory (to couple the geometric characteristics of the LGS to the sensor itself), and managing how to deal with the E2E simulations as the 2D Fourier method, commonly used, would not enhance its advantages. 

Until today we reached important results, which are documented in several reports, technical documents and proceedings (see Section~\ref{sec:ref}) presented during conferences (SPIE, AO4ELT, WFS workshops) and/or during internal meetings.
To resume:
\begin{itemize}
    \item The characteristics of the sensor are already defined thanks to a careful study that considered the configuration and the geometry of the ELT
    \item We conducted preliminary experiments in the laboratory using available materials, testing the response and the performance of the sensor and starting to contact the industry to upgrade the setup, making more suitable and compliant to the ELT expectations
    \item We concluded the tests at LAM laboratory and analyzed several configurations of the couple LGS $+$ Atmosphere, defining alignment procedures, data collection and analysis
    \item The simulation tool is already developed and we are working to speed up the operations and improve the compliance with the reality to run high-resolution simulations.
\end{itemize}

In the next future, to finalize the feasibility of the sensor, the project accounts for:
\begin{itemize}
    \item the final development of the simulation tool to estimate the expected sensibility and the response of the system (in the ELT perspective) and its usability to the whole group;
    \item the publications of the results, in particular their comparison with the ones obtained using different types of sensors, (pyramid and Shack-Hartmann WFS) and the evaluation of the better performance;
    \item the final implementation of the laboratory setup at INAF-Padova and publication of the tests made to learn the sensibility of the instrument and the definition of the common operations needed for the alignment of the whole system;
    \item the prototyping of the prism to draw attention to possible difficulties for the implementation and integration of the system;
    \item the definition of the design of an opto-mechanical system and possible realization of an LGS simulator with a turbulence generator in the context of the Laser Laboratory at INAF-Rome to verify the performance of the prototype
    \item the feasibility study for an experiment at the ESO instrument WLGSU and with the WHT AO module Canary.
\end{itemize}

Therefore, the next phases that have to be faced are sequential and their succession goes hand in hand with the knowledge of the sensor response and with the skills that the whole team acquires in a trade-off training process.
We wish to develop the  prototype soon and publish some exciting news once we will test it on sky.

\acknowledgments 
 We acknowledge the ADONI Laboratory for the support in the development of this project. 
 Funding to support the lab activities described in this paper came also from the INAF Progetto Premiale "Ottica Adattiva Made in Italy per i grandi telescopi del futuro".

\bibliography{main} 

\begin{thebibliography}{10}

\bibitem{Gilmozzi2007}
{Gilmozzi}, R. and {Spyromilio}, J., ``{The European Extremely Large Telescope
  (E-ELT)},'' {\em The Messenger}~{\bf 127} (Mar. 2007).

\bibitem{Johns2008}
{Johns}, M., ``{The Giant Magellan Telescope (GMT)},'' in [{\em Extremely Large
  Telescopes: Which Wavelengths? Retirement Symposium for Arne
  Ardeberg}{\nolinebreak\hspace{0.1em}]},  {\em SPIE Proceeding} {\bf 6986},  698603
  (Apr. 2008).

\bibitem{Ragazzoni2017}
{Ragazzoni}, R., {Portaluri}, E., {Viotto}, V., {Dima}, M., {Bergomi}, M.,
  {Biondi}, F., {Farinato}, J., {Carolo}, E., {Chinellato}, S., {Greggio}, D.,
  {Gullieuszik}, M., {Magrin}, D., {Marafatto}, L., and {Vassallo}, D.,
  ``{Ingot Laser Guide Stars Wavefront Sensing},'' {\em AO4ELT5 Proceeding}
  (Aug. 2017).

\bibitem{Ragazzoni2018}
{Ragazzoni}, R., {Greggio}, D., {Viotto}, V., {Di Filippo}, S., {Dima}, M.,
  {Farinato}, J., {Bergomi}, M., {Portaluri}, E., {Magrin}, D., {Marafatto},
  L., {Biondi}, F., {Carolo}, E., {Chinellato}, S., {Umbriaco}, G., and
  {Vassallo}, D., ``{Extending the pyramid WFS to LGSs: the INGOT WFS},'' in
  [{\em Adaptive Optics Systems VI}{\nolinebreak\hspace{0.1em}]},  {Close},
  L.~M., {Schreiber}, L., and {Schmidt}, D., eds., {\em Society of
  Photo-Optical Instrumentation Engineers (SPIE) Conference Series} {\bf
  10703},  107033Y (July 2018).

\bibitem{Ragazzoni2019}
{Ragazzoni}, R., Viotto, V., Portaluri, E., Bergomi, M., Greggio, D., {Di
  Filippo}, S., Radhakrishnan, K., Umbriaco, G., Dima, M., Magrin, D.,
  Farinato, J., Marafatto, L., Arcidiacono, C., and Biondi, F., ``Pupil plane
  wavefront sensing for extended and 3d sources,'' in [{\em
  AO4ELT6}{\nolinebreak\hspace{0.1em}]},  (2019).

\bibitem{Fried1995}
{Fried}, D.~L., ``{Focus anisoplanatism in the limit of infinitely many
  artificial-guide-star reference spots.},'' {\em Journal of the Optical
  Society of America A}~{\bf 12},  939--949 (May 1995).

\bibitem{Pfrommer2009}
{Pfrommer}, T., {Hickson}, P., and {She}, C.-Y., ``{A large-aperture sodium
  fluorescence lidar with very high resolution for mesopause dynamics and
  adaptive optics studies},'' {\em GRL}~{\bf 36},  L15831 (Aug. 2009).

\bibitem{Viotto2018}
{Viotto}, V., {Portaluri}, E., {Arcidiacono}, C., {Ragazzoni}, R., {Bergomi},
  M., {Di Filippo}, S., {Dima}, M., {Farinato}, J., {Greggio}, D., {Magrin},
  D., and {Marafatto}, L., ``{Dealing with the cigar: preliminary performance
  estimation of an INGOT WFS},'' in [{\em Adaptive Optics Systems
  VI}{\nolinebreak\hspace{0.1em}]},  {Close}, L.~M., {Schreiber}, L., and
  {Schmidt}, D., eds., {\em Society of Photo-Optical Instrumentation Engineers
  (SPIE) Conference Series} {\bf 10703},  107030V (July 2018).

\bibitem{Viotto2019}
Viotto, V., Portaluri, E., Arcidiacono, C., Bergomi, M., {Di Filippo}, S.,
  Greggio, D., Radhakrishnan, K., Dima, M., Farinato, J., Magrin, D.,
  Marafatto, L., and Ragazzoni, R., ``Ingot wavefront sensor: Simulation of
  pupil images,'' in [{\em AO4ELT6}{\nolinebreak\hspace{0.1em}]},  (2019).

\bibitem{Portaluri2019}
Portaluri, E., Viotto, V., Ragazzoni, R., Arcidiacono, C., Bergomi, M.,
  Greggio, D., Radhakrishnan, K., {Di Filippo}, S., Marafatto, L., Dima, M.,
  Biondi, F., Farinato, J., and Magrin, D., ``Ingot wfs for lgss: First results
  from simulations,'' in [{\em AO4ELT6}{\nolinebreak\hspace{0.1em}]},  (2019).

\bibitem{2020SPIEElisa}
{Portaluri}, E., {Viotto}, V., {Ragazzoni}, R., {Arcidiacono}, C., {Greggio},
  D., {Radhakrishnan Santhakumari}, K.~K., {Bergomi}, M., {Di Filippo}, S.,
  {Farinato}, J., and {Magrin}, D., ``{Evaluating the performance of an Ingot
  wavefront sensor for the ELT: good news from simulations},'' in [{\em Society
  of Photo-Optical Instrumentation Engineers (SPIE) Conference
  Series}{\nolinebreak\hspace{0.1em}]},  {\em Society of Photo-Optical
  Instrumentation Engineers (SPIE) Conference Series} {\bf 11448},  114483I
  (Dec. 2020).

\bibitem{2020SPIEKalyan}
{Radhakrishnan Santhakumari}, K.~K., {Greggio}, D., {Bergomi}, M., {Di
  Filippo}, S., {Viotto}, V., {Portaluri}, E., {Arcidiacono}, C., {Dima}, M.,
  {Lessio}, L., {Marafatto}, L., {Furieri}, T., {Bonora}, S., and {Ragazzoni},
  R., ``{Aligning and testing the ingot wavefront sensor in the lab},'' in
  [{\em Society of Photo-Optical Instrumentation Engineers (SPIE) Conference
  Series}{\nolinebreak\hspace{0.1em}]},  {\em Society of Photo-Optical
  Instrumentation Engineers (SPIE) Conference Series} {\bf 11448},  1144860
  (Dec. 2020).

\bibitem{DiFilippo2022}
{Di Filippo}, S., {Greggio}, D., {Bergomi}, M., {Radhakrishnan Santhakumari},
  K.~K., {Portaluri}, E., {Arcidiacono}, C., {Viotto}, V., {Ragazzoni}, R.,
  {Dima}, M., {Marafatto}, L., {Farinato}, J., and {Magrin}, D., ``{Laboratory
  testing of the Ingot WFS},'' in [{\em This
  conference}{\nolinebreak\hspace{0.1em}]},  {\em Society of Photo-Optical
  Instrumentation Engineers (SPIE) Conference Series} (July 2022).

\bibitem{Arcidiacono2020}
{Arcidiacono}, C., {Di Filippo}, S., {Greggio}, D., {Radhakrishnan
  Santakumari}, K.~K., {Portaluri}, E., {Bergomi}, M., {Viotto}, V., {Magrin},
  D., {Ragazzoni}, R., {Marafatto}, L., {Dima}, M., {Farinato}, J.,
  {Janin-Potiron}, P., {Fusco}, T., {Neichel}, B., {Fauvarque}, O., and
  {Schatz}, L., ``{Ingot wavefront sensor: from the Fourier End2End numerical
  simulation to the LOOPS test bench},'' in [{\em Society of Photo-Optical
  Instrumentation Engineers (SPIE) Conference
  Series}{\nolinebreak\hspace{0.1em}]},  {\em Society of Photo-Optical
  Instrumentation Engineers (SPIE) Conference Series} {\bf 11448},  1144868
  (Dec. 2020).

\bibitem{DiFilippo2019}
{Di Filippo}, S., Greggio, D., Bergomi, M., Radhakrishnan, K., Portaluri, E.,
  Viotto, V., Arcidiacono, C., Magrin, D., Marafatto, L., Dima, M., Ragazzoni,
  R., Janin-Portirond, P., Schatz, L., Neichel, B., Fauvarque, O., and Fusco,
  T., ``Ingot wavefront sensor: from the optical design to a preliminary
  laboratory test,'' in [{\em AO4ELT6}{\nolinebreak\hspace{0.1em}]},  (2019).

\end{thebibliography}
\label{sec:ref}
\end{document}